\documentclass[10pt,letterpaper]{article}
\usepackage{opex3}
\usepackage{amsmath,amssymb}
\usepackage{mathrsfs}
\usepackage{graphicx}% Include figure files
\usepackage{dcolumn}% Align table columns on decimal point
\usepackage{bm}% bold math

\begin{document}

\title{Van der Waals enhancement of optical atom potentials via resonant coupling to surface polaritons}

\author{Joseph Kerckhoff and Hideo Mabuchi}
\address{Edward L.\ Ginzton Laboratory, Stanford University, Stanford CA 94305, USA}
\email{jkerc@stanford.edu}

\date{\today}

\begin{abstract}
Contemporary experiments in cavity quantum electrodynamics (cavity QED) with gas-phase neutral atoms rely increasingly on laser cooling and optical, magneto-optical or magnetostatic trapping methods to provide atomic localization with sub-micron uncertainty. Difficult to achieve in free space, this goal is further frustrated by atom-surface interactions if the desired atomic placement approaches within several hundred nanometers of a solid surface, as can be the case in setups incorporating monolithic dielectric optical resonators such as microspheres, microtoroids, microdisks or photonic crystal defect cavities. Typically in such scenarios, the smallest atom-surface separation at which the van der Waals interaction can be neglected is taken to be the optimal localization point for associated trapping schemes, but this sort of conservative strategy generally compromises the achievable cavity QED coupling strength. Here we suggest a new approach to the design of optical dipole traps for atom confinement near surfaces that exploits strong surface interactions, rather than avoiding them, and present the results of a numerical study based on $^{39}$K atoms and indium tin oxide (ITO). Our theoretical framework points to the possibility of utilizing nanopatterning methods to engineer novel modifications of atom-surface interactions.
\end{abstract}


\section{Introduction}

Over the past decade there have been numerous experimental and theoretical studies in atomic, molecular and optical physics~\cite{Folm00,Lev03,Wang05,Ghan06,Treutlein07,Purd08} and in quantum information science~\cite{Mabuchi01,Trup07,Daya08} focusing on the interaction of gas-phase neutral atoms with static or electromagnetic fields projected by planar micro- or nanofabricated structures. The tight dimensional control that can be achieved via modern lithographic techniques generally enables new regimes of performance and scalability in integrated atomic devices, which hopefully can be used to further the success of recent technology breakthroughs such as chip-scale atomic clocks~\cite{Schwindt04} and magnetometers~\cite{Knappe05}.

In many such scenarios---in particular those exploiting strong coupling of trapped atoms with optical fields confined in photonic waveguides or resonators---robust and accurate localization of atoms near solid surfaces emerges as a critical requirement. Indeed, the prospect of tight control over atomic ensembles alone has driven much of the development of integrated atomic platforms. Often atomic localization is achieved by ``scaling down'' trapping techniques first developed in free-space, as in the manipulation of neutral atoms by magnetic field gradients created by current-carrying microwires~\cite{Folm00} or by optical dipole forces induced by the evanescent tails of confined laser fields~\cite{Mabuchi94}. Robust and conceptually simple, these techniques typically improve in strength and efficiency as the atoms are brought closer to the defining structure(s). However, bulk surface effects become dominant at short distances. For small values of the atom-surface separation $r$, energy potentials for optical dipole traps and magnetostatic microwire traps are roughly proportional to $1-r/r_0$ and $r^{-1}$, respectively, whereas van der Waals potentials have a much stronger variation $\sim r^{-3}$. Hence many integrated atomic systems are expected to exist in an awkward regime where atoms are kept close to trap-defining structures to achieve tight potentials, but not so close that van der Waals effects become problematic. In the context of cavity quantum electrodynamics (cavity QED) with monolithic dielectric resonators this can be a particularly significant compromise, as it generally precludes realization of the strongest possible atom-photon coupling parameters, which would require stable placement of atoms nearly in contact with the resonator surface. For this reason and others~\cite{Henk05}, van der Waals interactions can thus become a limiting factor in the achievable performance of a wide range of integrated atomic devices, and we are motivated to investigate new approaches to atom manipulation that can compete with the strong onset of common $r^{-3}$ surface effects. It may be anticipated that such approaches will rely on the physics particular to atoms in the near field of bulk structures, rather than the reapplication of techniques developed {\it in vacuo}.

A reasonable first step in this direction could be to explore the design of trapping schemes that attempt to exploit serendipitous aspects of the atom-surface interaction. For example, the demonstration of a \emph{positive} van der Waals energy shift of a particular excited state of Cs over an oriented sapphire crystal that is orders of magnitude larger than typical, attractive van der Waals potentials~\cite{Ducloy99,Failache03,Ducloy02} suggests a possible means of surface-specific manipulation of atoms. If the van der Waals, repulsive character of an excited atomic state may be accessed appropriately in the near-field of a surface, it would provide a valuable tool for the efficient localization of atoms very close to structures of interest. The anomalous van der Waals perturbation demonstrated in \cite{Failache03} arises from the off-resonant coupling of an atomic transition to a surface polariton resonance in the sapphire crystal, suggesting that analogous and potentially improved phenomena could be accessed in alternative systems chosen or even created using modern methods and insights from materials science and nanofabrication.

In this article we describe and analyze (through numerical simulations) a scheme for using off-resonant coupling of gas-phase atoms to surface polariton resonances in a surface near field to produced van der Waals-\emph{enhanced} optical potentials. In section \ref{interactions} we summarize previous work that connects atom-surface van der Waals effects to a perturbative interaction with surface field modes. Then in section \ref{mirrors} we describe a general approach to harnessing these effects with laser fields to produce surface enhanced optical potentials. Section \ref{example} illustrates this approach with simulations of specific systems based on $^{39}$K and a planar indium tin oxide (ITO) surface. And in concluding, we mention some possible paths toward engineering general, useful van der Waals interactions.

\section{Atom-surface interactions}\label{interactions}

\noindent QED interactions between atoms and/or bulk materials have been analyzed in many contexts over almost 80 years (e.g. \cite{London30,Dzy61}), but this section will largely summarize relevant results in work by Wylie and Sipe \cite{Wylie84,Wylie85,Sipe81} (whose perspective we adopt throughout the article) as they naturally highlight the importance of surface excitations in producing anomalous van der Waals effects.

An atom near a macroscopic structure experiences effects often dominated by the perturbing interaction of the atomic dipole and radiation field
\begin{equation}
H_i = -\vec{\mu}\cdot\vec{D},
\end{equation}
where $\vec{\mu}$ is the atomic dipole operator and $\vec{D}$ is the transverse displacement field operator at the atom's position.  Thus, just as atomic level shifts and the spontaneous decay of excited states arise from interactions with radiation modes in a vacuum, van der Waals effects are produced by similar physical processes, with contributions from a near field, planar surface.  For example, in a purely semi-classical description, an excited atomic dipole near an interface will have a reaction to the reflected portion of its own induced radiation field.  As the amplitude of the reflected field at the atom's position increases, so do these reactive, surface-induced effects.

Without having to specify any geometry or optical properties of the system yet, the energy shift of atomic level $a$ may be approximated by second order perturbation theory
\begin{equation}\label{2PT}
\delta E_a = -\hbar^{-1}\mathbf{P}\sum_{B,N,n}p(B)\frac{D^{BN}_\alpha D^{NB}_\beta\mu^{an}_\alpha\mu^{na}_\beta}{(\omega_N-\omega_B)+(\omega_n-\omega_a)}
\end{equation}
where $\mathbf{P}$ indicates the principal part, the summation is over field states $B$ and $N$ and atomic states $n$, $\hbar\omega_i$ is the energy eigenvalue for un-perturbed field or atomic state $i$, $p(B)$ denotes an assumed thermal distribution of the field, and the summation over vector components $\alpha$ and $\beta$ is implied.  Also, the transition rate from atomic state $a$ to $n$ is given by Fermi's golden rule
\begin{equation}\label{Fgr}
R_{na} = \frac{2\pi}{\hbar}\sum_{B,N}p(B)|\langle n,N|\vec{\mu}\cdot\vec{D}|a,B\rangle|^2\delta((\omega_N-\omega_B)+(\omega_m-\omega_a)).
\end{equation}
We introduce the correlation functions
\begin{eqnarray}
\tilde{G}_{\alpha\beta}(t)& = & i\hbar^{-1}\langle[D_\alpha(t),D_\beta(0)]\rangle\Theta(t)\nonumber\\
\tilde{\alpha}_{\alpha\beta}^a(t) & = & i\hbar^{-1}\langle a|[\mu_\alpha(t),\mu_\beta(0)]|a\rangle\Theta(t),\label{correlation}
\end{eqnarray}
where $D_\alpha(t)$ and $\mu_\alpha(t)$ are interaction picture operators with respect to $H_i$ and $\Theta(t)$ is the Heaviside step function, and also specify their Fourier transforms: $G_{\alpha\beta}(\omega)$ and $\alpha_{\alpha\beta}^a(\omega)$.  Eqs. \eqref{correlation} describe the linear response of operator expectation $\langle X(t)\rangle$, given a perturbation proportional to $X(0)$.  And, from linear response theory, $G_{\alpha\beta}(\omega)$ and $\alpha_{\alpha\beta}^a(\omega)$ are the generalized susceptibilities for the field and the atom.  In particular, $G_{\alpha\beta}(\omega)$ may be identified with the expected displacement field induced by an oscillating \emph{classical} dipole, and $\alpha_{\alpha\beta}^a(\omega)$ the polarizability of the $a^\text{th}$ atomic state by the field.  The fluctuation-disspation theorem, a zero temperature approximation (appropriate for optical frequencies at room temperature), and the analytic properties of the susceptibilities allow Eqs. \eqref{2PT} and \eqref{Fgr} to be written
\begin{eqnarray}
\delta E_a & = & \delta E_a^{vf} + \delta E_a^{r}\nonumber\\
& = & -\frac{\hbar}{2\pi}\int_0^\infty d\zeta G_{\alpha\beta}(i\zeta)\alpha_{\alpha\beta}^a(i\zeta)\nonumber\\
&&-\sum_n\mu_\alpha^{an}\mu_\beta^{na}\text{Re} G_{\alpha\beta}(\omega_{an})\Theta(\omega_{an})\label{dE}\\
R_{na} & = & \frac{2}{\hbar}\mu_\alpha^{an}\mu_\beta^{na}\text{Im}G_{\alpha\beta}(\omega_{an})\Theta(\omega_{an})\label{R}
\end{eqnarray}
where $\omega_{an} = \omega_a-\omega_n$.

One of the advantages of this approach is that the contribution to $G_{\alpha\beta}(\omega)$ from an interface may be simply added that of free space by the superposition principle, and therefore the surface effects can be considered separately from the (formally divergent) free space effects.  From here on, we make this separation and only consider the effects from field modes emanating from the surface.  The first term on the RHS of Eq. \eqref{dE}, $\delta E_a^{vf}$, may be associated with the polarization energy of the atom due to quantum fluctuations in these surface modes, in analogy to the Lamb shift in free space. As the term integrates the susceptibilities over imaginary frequencies, this shift is always real, usually negative~\footnote{Each $n$-level's contribution to $\delta E_a^{vf}$ carries the sign of $\omega_{na}$, as is suggested from Eq. \eqref{polarizability}.  Thus for downward transitions, $\delta E_a^{vf}$  will be positive, but the net shift will often be negative because $\delta E_a^{r}\sim-2|\delta E_a^{vf}|$ for any interface that is moderately reflective at $|\omega_{na}|$ (see Eq. \ref{dE_nf}).}, and generally insensitive to resonances in the environment, giving rise to the familiar, attractive van der Waals potential for ground state atoms.  This is in contrast to the last term in Eq. \eqref{dE}, $\delta E_a^{r}$, which arises from the residue of the principal part integration in Eq. \eqref{2PT}.  This term has none of the analytic constraints of the first term: it may be either positive or negative, and is highly sensitive to environmental resonances near $\omega_{an}$.  Due to the zero temperature field assumption, this residue exists only for \emph{excited} atomic states, which may be degenerate with other states of the atom plus field.  As it originates from global energy degeneracies, this term admits a semi-classical physical interpretation of the energy associated with an atomic dipole interacting with the \emph{in phase} portion of its own, expected reflected field.  Similarly, Eq.~\eqref{R} may be viewed in analogy to the dephasing rate of an excited dipole from the interaction with the \emph{out of phase} portion of the expected reflected field.

\begin{figure}
\begin{center}
\includegraphics[width=2.5in]{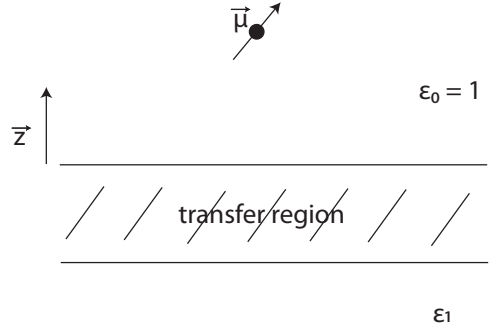}
%\vspace{-3mm}
\caption[figure1]{\label{fig:dipole_interface} The assumed geometry of an oriented atomic dipole in vacuum at height $z$ above a planar, transfer region interface that separates an underlying bulk substrate with dielectric constant $\epsilon_1$ from the vacuum.  Adapted from \cite{Sipe81}.}
\end{center}\vspace{-0.3in}
\end{figure}

The atomic polarizability can be shown to be
\begin{equation}\label{polarizability}
\alpha_{\alpha\beta}^a(\omega) = \frac2\hbar \sum_n\frac{\omega_{na}\mu_\alpha^{an}\mu_\beta^{na}}{\omega_{na}^2-\omega^2},
\end{equation}
while the field susceptibility may be constructed from a plane wave decomposition of the reflected electromagnetic field produced by an oscillating dipole above a particular surface.  To specify now some general aspects of the system, the interface is assumed to be translationally invariant in the $(xy)$ plane (see figure \ref{fig:dipole_interface}) which gives
\begin{equation}\label{G}\begin{split}
G_{\alpha\beta}(\omega) = \frac{i\tilde{\omega}^2}{2\pi}\int\frac{d\vec{\kappa}}{W_0}(\hat{s}_\alpha\hat{s}_\beta R^s+\hat{p}^{0+}_\alpha\hat{p}^{0-}_\beta R^p)\exp{(2iW_0z)}
\end{split}\end{equation}
where $\tilde{\omega} = \omega c^{-1}$, $\vec{\kappa}$ and $W_0$ are the components of the vacuum wave vector at the transfer region-vacuum interface perpendicular and parallel to $\vec{z}$, respectively, $\hat{s}$ and $\hat{p}^{0\pm}$ are the s- and p-polarized electric field vectors in the vacuum for upward $(+)$ and downward $(-)$ propagating fields, $R^{s,p}$ are the Fresnel coefficients of the transfer region for the respective polarizations, and $z$ is the height of the atom above the surface of the transfer region.

Although surface interactions are typically considered in the near field limit, which leads to a dramatic simplification of $G_{\alpha\beta}(\omega)$, Eq. \eqref{G} is especially useful for engineering considerations, as it directly connects the effects on the atom by the surface with an intuitive picture of the dipole interacting with all possible, reflected surface modes at frequency $\omega$.  In particular, if the Fresnel coefficients are large over a large range of $\vec{\kappa}$ at an atomic transition frequency $\omega_{an}$, the corresponding effects described by Eq. \eqref{R} and the last term of Eq. \eqref{dE} will be likewise large and carry the sign of the integrated Fresnel coefficients.

The common case of an atom above a bulk substrate (no transfer region) with separations $z\ll\tilde{\omega}_{na}$ may be considered by taking the near-field limit ($c\rightarrow\infty$)
in Eq. \eqref{G}.  In this limit, the Fresnel coefficients become $
\vec{\kappa}$-independent, and the integral may be simply evaluated:
\begin{eqnarray}
G_{zz}(\omega)&=&2G_{xx}(\omega)=2G_{yy}(\omega)\nonumber\\\label{G_nf}
&=& \frac{1}{4z^3}\frac{\epsilon_1(\omega)-1}{\epsilon_1(\omega)+1}
\end{eqnarray}
where the final ``image factor'' is the familiar, ``quasi-static,'' p-polarized Fresnel coefficient of the interface (and all other $G_{\alpha\beta}(\omega)$ components are $0$).  In this limit, all surface modes coupling to the atom are evanescent (Re$W_0=0$), and therefore bound to the surface.  Resonant surface excitations in \emph{all} $\vec{\kappa}$ modes are excited by the atom as $\epsilon_1(\omega_{an})\rightarrow-1$, leading to a divergent response in $\delta E_a^{r}$ and $R_{na}$.  Significantly, an anomalously large, \emph{positive}, van der Waals-like energy shift of excited atomic states may be realized by off-resonant coupling to these surface excitations in the near field as $\epsilon(\omega_{an})+1\rightarrow0^+$.  An intuitive explanation for this effect is that the atomic dipole may interact with a near field image dipole (without retardation) that because of material resonances is both larger than and $\sim180^\circ$-shifted from the image dipole in an ordinary reflector.

For example, consider a Drude metal with the dielectric function
\begin{equation}\label{Drude}
\epsilon_D(\omega) = 1-\frac{\omega_p^2}{\omega(\omega+i\gamma)}
\end{equation}
where $\omega_p$ is the metal's plasma frequency and $\gamma$ the electron collision rate.  Using $\epsilon_D$, the real part of the image factor in Eq. \eqref{G_nf} reaches a minimum of $\frac{-1}{2}(\frac{\sqrt2\gamma}{\omega_p}+(\frac{\gamma}{\omega_p})^2)^{-1}\equiv I_{min}$ when $\omega=(\frac{\omega_p^2}{2}+\frac{\gamma\omega_p}{\sqrt{2}})^{1/2}\equiv\omega_{min}$.  Compared to $1$, image factor for a perfect conductor, this minimum coefficient has the opposite sign and is far larger in magnitude when $\omega_p\gg\gamma$.  The corresponding imaginary part of the minimum image factor is $-I_{min}(1+\sqrt{2}\frac{\gamma}{\omega_p})^{1/2}$, which notably also diverges for a ``dissipation free'' metal, $\gamma\rightarrow0$.  An excited atom with a \emph{downward} transition frequency near $\omega_{min}$ in the near field of this Drude metal will experience a positive energy shift proportional to $z^{-3}$ that will dominate $\delta E_a^{vf}$ when $\omega_p\gg\gamma$.  Also, the spontaneous emission rate of that particular transition will likewise be enhanced by a factor of order $\hbar^{-1}\delta E_a^r$, proportional to $z^{-3}$.

\section{Van der Waals atom mirrors}\label{mirrors}

The flexibility of $\delta E_a^{r}$ suggests that a resonant material interface may be used to engineer the energy potential of an atom in its near field.  However,  to do so we are required to find a suitable means of accessing these \emph{excited} state effects, while keeping the atomic heating from spontaneous emission manageable.

\begin{figure}
\begin{center}
\includegraphics[width=4in]{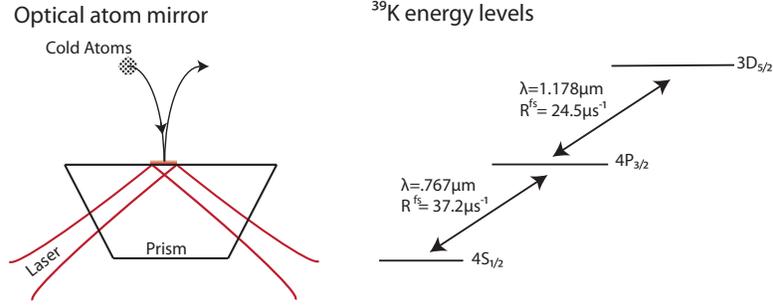}
%\vspace{-3mm}
\caption[figure1]{\label{fig:atom_mirror} Left: diagram of a typical optical atom mirror, as described in section \ref{mirrors}.  Right: the $^{39}$K fine structure energy levels relevant to the example in section \ref{example}. }
\end{center}\vspace{-0.3in}
\end{figure}

Previously used to measure the attractive van der Waals potential for alkali atoms near transparent dielectrics \cite{Landragin96}, optical atom mirrors could provide a simple platform to study and begin to exploit anomalous interactions of atoms and surface polaritons.  In a traditional optical atom mirror, an evanescent, coherent field is produced by total internal reflection of a laser beam from a planar, vacuum-dielectric interface (see figure \ref{fig:atom_mirror}).  If the frequency of the laser is far blue-detuned from a ground state transition, an atom approaching the evanescent field experiences a repulsive optical dipole force proportional to the intensity of the field:
\begin{equation}\label{atom_mirror_pot}
U_{d} \approx \frac{\hbar\Omega_0^2}{4\Delta}e^{-2\kappa_zz}
\end{equation}
where $\Omega_0$ is the Rabi oscillation frequency the atom would undergo at the dielectric surface if the laser was on resonance, $\Delta$ is the detuning of the field from the transition frequency, and $\kappa_z$ is the field amplitude decay rate (typically of order the angular wave number of the laser).  For sufficiently large  $U_d$, this optical potential may repel incident, cold thermal atoms (e.g. falling in vacuum under the influence of gravity) with minimal heating from spontaneous emission.  The inclusion of van der Waals effects adds an attractive, $\sim z^{-3}$, potential associated with the red-shifting of the atomic ground state (as measured in \cite{Landragin96}), but also modifies the laser detuning as both ground and excited states are perturbed.  These surface-induced modifications to the optical potential may be used to produce van der Waals-\emph{enhanced} optical potentials, where the repulsive energy is derived from the laser fields, but the spatial variation of the potential carries the sharp character of the surface interactions: as the van der Waals, $\sim z^{-3}$, modifications occur predominantly in the near field where the un-perturbed $U_d$ is roughly linear, the enhancement will be sharply peaked.  As a simple example, if the excited state experiences a very large, positive van der Waals shift from coupling to resonant surface excitations, the field detuning will decrease faster than the red-shifting of the ground state.  In the perturbative regime, $\Delta\gg\delta E_i$, the van der Waals contributions to the potential are
\begin{equation}\label{vW_atom_mirror}
U_{vdW}\sim\frac{\Omega_0^2}{4\Delta^2}\delta E_e^{r}+\delta E_g^{vf},
\end{equation}
and so an appropriately large and positive $\delta E_e^{r}$ will provide a steep ($\sim z^{-3}$) and repulsive $U_{vdW}$.  Also note that the probe need not be evanescent in this case to achieve near field repulsion.  However, it is important to recognize that this system will also experience an enhanced rate of spontaneous emission in the near field $\approx \frac{\Omega_0^2}{4\Delta^2}R_{e}$, which will be comparable to the optical potential enhancement if $\delta E_{e}\sim \hbar R_e$, as suggested in the end of section \ref{interactions}.  Although this necessarily increases the heating rate, surface-induced spontaneous emission enhancement can sharply redistribute atomic populations in the near field \cite{Ducloy02} to advantageous effect.  For example, enhanced emission may ``turn off'' target optical dipole transitions and thus quench dipole forces.  If a ground state atom experiences balanced attractive and repulsive optical dipole forces at moderate atom-surface separations \cite{Mabuchi94}, greatly enhanced spontaneous emission on the ``attractive'' dipole transition can produce a net repulsive force in the surface near field, producing a sharp barrier and preventing collisions with the surface.   As this general approach is particularly valid when spontaneous emission enhancement dominants any energy level shifts, its application may be particularly useful when the stability of atomic transitions is required.  Moreover, in comparison, to other ``repulsive'' van der Waals interactions that rely on the reflection of atomic wavepackacts from steeply attractive van der Waals potentials \cite{Pasquini04}, both approaches outlined above are designed to manipulate the classical center of mass, and thus may be performed with atoms with relatively small de Broglie wavelengths.

More accurate and general optical forces for thermal atoms may be predicted from the steady state solution to the atomic master equation ($\hbar = 1$) \cite{API}:
\begin{equation}\begin{split}\label{OBE}
\dot{\rho} = \mathcal{L}(\rho)\\
\mathcal{L}(\rho) = i\left[\rho,H\right] + \sum_{a,n}\mathcal{D}_{na}(\rho)\\
H = \frac12\left(\sum_{a,n} \Omega_{an}\sigma_{na}-\sum_{a} \Delta_a\sigma_{aa}\right)+ \text{h.c.}\\
\mathcal{D}_{an}(\rho) = R^{tot}_{na}\left(\sigma_{na}\rho\sigma_{na}^\dag-\frac12\left\{\sigma_{na}^\dag\sigma_{na},\rho\right\}\right)
\end{split}\end{equation}
where $\rho$ is the atomic state density matrix, $\sigma_{na}$ is the atomic lowering operator $|n\rangle\langle a|$, $\Omega_{an}$ is the Rabi frequency induced between states $a$ and $n$, $\Delta_a$ is the energy of the laser photons near-resonant with the transition of a ground state atom to level $a$ minus the energy of state $a$, $R^{tot}_{na}$ is the total decay rate of state $a$ to $n$, and ($\{\cdot,\cdot\}$) $[\cdot,\cdot]$ is the (anti-)commutator.  In the cases considered below, the nullspace of the superoperator $\mathcal{L}$ contains a unique atomic steady state $\rho_{ss}$, to which an arbitrary initial atomic state will decay in time of order $\text{max}\{R_{na}^{-1}\}$.  Typically, the internal state of the atom relaxes much faster than the external parameters change (i.e. the velocity of the atom times the spatial derivative of the coefficients in Eq. \eqref{OBE} is much less than $R_{na}$).  In such regimes, use of $\rho_{ss}$ to estimate the force on the atom as a function position may be justified.  To associate an internal atomic state to a force on the center of mass (CM) motion of the atom, we adopt a semi-classical approximation that assumes quantized internal dynamics (Eq. \eqref{OBE}), but classical CM dynamics.  In particular, as the expected force on the center of mass of an atom by Ehrenfest's theorem is $-\text{Tr}([\nabla_{\mathbf{R}_{CM}},H]\rho)$ (where $\mathbf{R}_{CM}$ is the center of mass position operator), declaring the center of mass a ``classical'' variable, $r_{CM}$, allows the construction of an effective potential for the CM motional dynamics
\begin{equation}\label{Ueff}
U_{eff}(r_{CM})=\int_\infty^{r_{CM}}dr \text{Tr}(\frac{\delta H(r)}{\delta r}\rho_{ss}(r)).
\end{equation}
This description should be most accurate at capturing the essential contributions to the mean force on slowly moving atoms with small de Broglie wavelengths from \emph{reactive} (``dipole'') interactions with the laser fields \cite{API}.  Similarly, it is straightforward to show that the steady state heating rate of the atom due to isotropic spontaneous emission is
\begin{equation}\label{Heat}
\frac{\hbar^2}{3 c^2k_Bm}\sum_{a,n}\omega_{na}^2R_{na}^{tot}\text{Tr}(\sigma_{aa}\rho_{ss}(r_{CM}))
\end{equation}
where $m$ is the mass of the atom.

\section{Example}\label{example}
\subsection{ $^{39}$K and ITO}\label{atom_mirror}
	
	As described in section \ref{interactions} and appendix \ref{calculation}, resonant van der Waals effects are a product of matched, high quality surface polariton and strong optical dipole transition resonances.  As is suggested by Eq. \ref{dipole_strength}, dipole transition moments tend to increase rapidly with wavelength ($\sim|\omega_{na}|^{-3}$).  Similarly, compared to plasmonic metals like Au and Ag, relatively stiff materials like sapphire, Y$_2$O$_3$, and SiC tend to feature high quality surface phonon-polariton excitations near crystal lattice vibrational modes in the far infra-red ($<$1000cm$^{-1}$) \cite{Rakic98,Saltiel06,Palik}.  Thus, strong resonant interactions with common bulk materials are likely to occur in multiply-excited states of alkali atoms, which have strong, $\sim10\mu$m wavelength transitions \cite{Heavens61,Lindgard77}.  However, working with these highly excited atomic states is often very difficult.  For example, although the resonant van der Waals interaction of the Cs $6D_{3/2}\rightarrow7P_{1/2}$ (12-$\mu$m) transition and oriented crystal sapphire is very large \cite{Failache03}, an even partially excited $6D_{3/2}$ state will populate so many lower-lying levels through spontaneous emission \cite{Heavens61} that construction of an enhanced optical potential is difficult both to model and achieve.

However, there are combinations of common materials and convenient alkali states that admit a significant interaction without large complications from the multi-level atomic structure.  The relatively strong, $3D_{5/2}\rightarrow4P_{3/2}$ 1.178$\mu$m transition in neutral $^{39}$K \cite{Corliss79} is slightly to the blue of a surface polariton resonance recently identified in indium tin oxide (ITO) \cite{Rhodes08,Franzen08}, which suggests that the $3D_{5/2}$ state may experience a net \emph{positive} van der Waals energy shift in the vicinity of ITO.  Due to dipole transition selection rules, $3D_{5/2}$ states spontaneously decay only to $4P_{3/2}$, and $4P_{3/2}$ decays only to the $4S_{1/2}$ ground states \cite{Heavens61}.  Thus optically probing the $4P_{3/2}\leftrightarrow4S_{1/2}$ and  $3D_{5/2}\leftrightarrow4P_{3/2}$ transitions forms a closed system between these fine-structure states.  Moreover, cycling transitions between the $|4S_{1/2}, F=2,m_F=\pm2\rangle\leftrightarrow|4P_{3/2},F=3,m_F=\pm3\rangle\leftrightarrow|3D_{5/2},F=4,m_F=\pm4\rangle$ hyperfine magnetic sublevels are possible ($^{39}$K has a 3/2 nuclear spin \cite{Arimondo77}).  Calculation of the van der Waals effects on these fine structure states by a near field planar ITO film is done in appendix \ref{calculation} and presented in table \ref{vW_table}.  The results anticipate a net positive energy shift of the $3D_{5/2}$ state, as well as a significantly enhanced spontaneous emission rate on the 1.178$\mu$m transition.

\begin{figure}
\begin{center}
\includegraphics[width=4.5in]{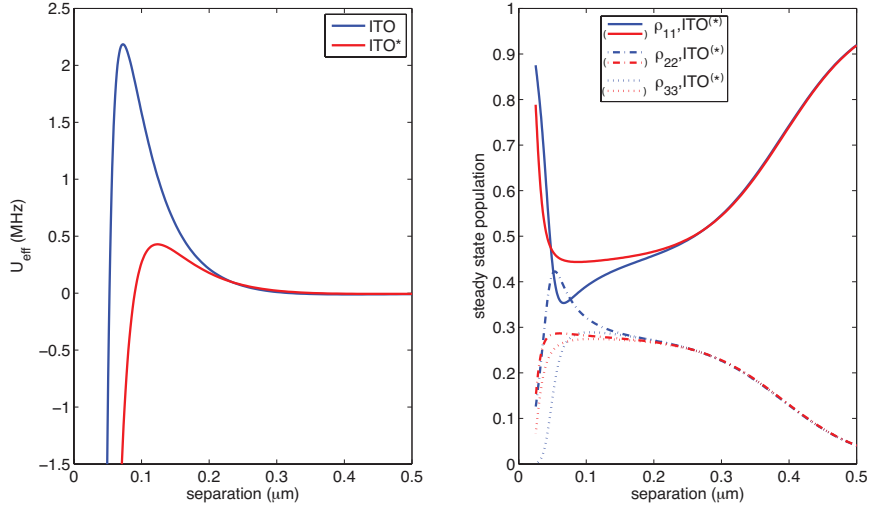}
%\vspace{-3mm}
\caption[figure1]{\label{fig:K_ITO} Left: effective, steady state optical potentials as a function of $^{39}$K-surface separation for both ITO and (dispersionless) ITO* materials with $\{\Omega_1,\Delta_1,\kappa_1,\Omega_2,\Delta_2\}/2\pi = \{100\text{MHz},50\text{MHz},(767\text{nm})^{-1},100\text{MHz},0\text{MHz}\}$ and atomic parameters from tables \ref{vW_components} and \ref{vW_table}.  Right: steady state atomic level populations for the same.}
\end{center}\vspace{-0.3in}
\end{figure}

In figure \ref{fig:K_ITO} we plot the steady state energy expectation (Eq. \eqref{Ueff}) of a van der Waals-enhanced optical atom mirror for $^{39}$K above an ITO interface as a function of atom-surface separation.  Employing a probe configuration that exploits both the van der Waals energy shifts and enhanced emission to produce a sharply peaked potential in the surface near field, the $4P_{3/2}\leftrightarrow4S_{1/2}$ 766.7nm transition is driven with maximum Rabi frequency, detuning, and field amplitude decay $\{\Omega_1,\Delta_1,\kappa_1\}/2\pi = \{100\text{MHz},50\text{MHz},(767\text{nm})^{-1}\}$ and the $3D_{5/2}\leftrightarrow4P_{3/2}$ 1.178$\mu$m transition is driven with $\{\Omega_2,\Delta_2\}/2\pi = \{100\text{MHz},0\text{MHz}\}$.  As ITO is quite lossy at 1.178$\mu$m (see Eq. \ref{Drude_ITO}), the second transition is assumed to be propagating in vacuum ($\kappa_2 = 0$) rather than evanescent.  The free-space, spontaneous decay rates of these levels are listed in table \ref{vW_components} and the additive enhancement of spontaneous emission rates and energy level shifts from van der Waals interactions are listed in table \ref{vW_table}.  To highlight the significance of the ITO polariton resonance, figure \ref{fig:K_ITO} depicts the effective, steady state optical potential for a $^{39}$K atom both in the near field of an ITO surface and in the near field of a similar, but dispersionless dielectric ``ITO*'' (see appendix \ref{calculation}).  The ITO potential exhibits a greatly enhanced potential barrier over the dispersionless surface, capable of deflecting cold atoms falling from mesoscopic heights.

\begin{figure}
\begin{center}
\includegraphics[width=4.5in]{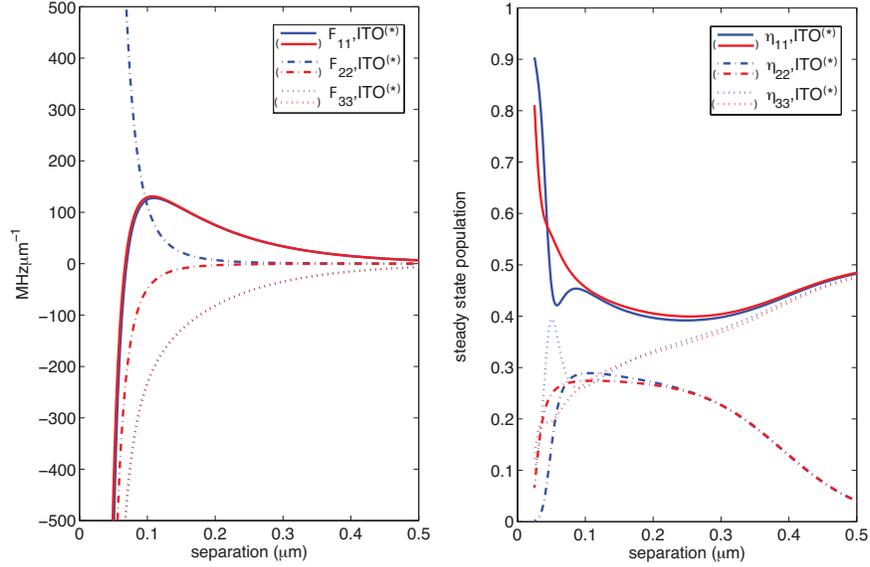}
%\vspace{-3mm}
\caption[figure1]{\label{fig:Force_ss} Left: eigenvalues of the force operator as a function of separation for the ITO and ITO* systems depicted in firgure \ref{fig:K_ITO}.  Right: steady state atomic populations in the (position-dependent) force eigenbasis for the same.}
\end{center}\vspace{-0.3in}
\end{figure}

We begin to understand the origins of the enhanced potential by diagonalizing the ``force operator'' $F = -\frac{\delta H}{\delta r}$ for the two potentials as a function of position and considering its eigenvalues.  As plotted in figure \ref{fig:Force_ss}, in its eigenbasis, the diagonal elements of $F$ from the far field up to about $.2\mu$m suggest that the two systems feature identical repulsive and attractive optical dipole force eigenstates, as well as a state that experiences no force.  In the near field, each force eigenstate attains a van der Waals-like character, with the most significant difference between the ITO and ITO* systems emerging from the initially force-less state, due to the difference in $\Delta E_{3D_{5/2}}$ calculated for the two surfaces (table \ref{vW_table}): the polariton resonance in ITO produces a positive $\Delta E_{3D_{5/2}}$ and thus a repulsive $F_{22}$, while $F_{22}$ for the dispersionless ITO* has a familiar, attractive, van der Waals character.  The total expected, steady state forces, and thus the effective potentials depicted in figure \ref{fig:K_ITO}, may be constructed from these position-dependent force eigenvalues and the diagonal elements of the steady state atomic density matrix in the force eigenbases, $\eta_{ii}$.  The right half of figure \ref{fig:Force_ss} depicts the steady state $\eta$ diagonal elements.  As to be expected, both the ITO effective potential in figure  \ref{fig:K_ITO} and the population in the van der Waals-repulsed state $\eta_{22}$ in figure \ref{fig:Force_ss} begin to decrease with decreasing separation at about the same point ($\sim80$nm).  It may appear from figure \ref{fig:Force_ss} that the enhancement of the optical forces in the ITO system is due to the near-field repulsive $F_{22}$ eigenvalue.  While this is largely true, the close correspondence of $\eta_{11}$ in the two material systems is a coincidence caused by the large enhanced decay of the $3D_{5/2}$ state in the ITO system.  If this decay enhancement is artificially removed in the ITO system, ITO's $\eta_{11}$ is significantly smaller over the 60nm-250nm separation range, where $F_{11}$ is still repulsive.  In this respect, the optical forces in the ITO near field can be considered van der Waals-enhanced by both  $\Delta E_{3D_{5/2}}$ and $R_{4P_{3/2}\leftarrow3D_{5/2}}$.

We conclude this subsection with a mention of the optical powers required to achieve the drives in the above example.  It is informative to calculate the saturation intensity for an $a\rightarrow n$ electric dipole transition \cite{SteckNotes}
\begin{equation}\label{Isat}
I_{sat} = \frac{\hbar R_{na}^{fs}\omega_{na}^3}{12\pi c^2}\frac{2J_a+1}{2J_n+1}.
\end{equation}
Using the data in table \ref{vW_components}, the saturation intensities of the transitions we are driving are $I_{sat,S\leftrightarrow P} = 3.4$mWcm$^{-2}$ and $I_{sat,P\leftrightarrow D} = .47$mWcm$^{-2}$.  Using $I_{sat}$, optical intensity may be related to Rabi frequency by $I = 2I_{sat}(\Omega_{na}/R_{na}^{fs})^2$.  Thus, $\Omega_1 = 2\pi\times100$MHz requires 1.9Wcm$^{-2}$ of optical power at 767nm and $\Omega_2 = 2\pi\times100$MHz requires 618mWcm$^{-2}$ at 1.178$\mu$m.  Although these intensities are non-trivial, restricting the illumination region to 1mm$^{-2}$ should bring the power requirements comfortably within the range of commercial diode lasers.

\subsection{Quantum trajectory simulations}\label{qtraj}

\begin{figure}
\begin{center}
\includegraphics[width=4.5in]{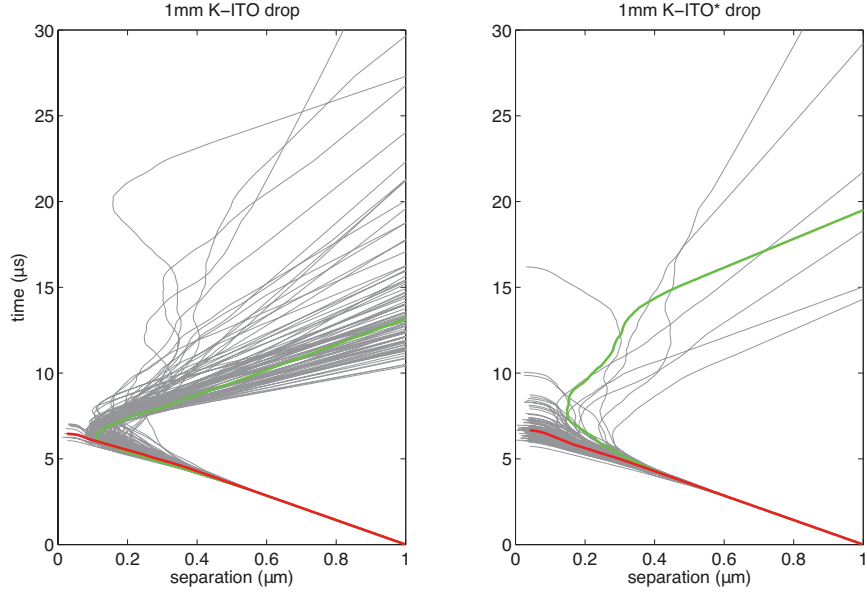}
%\vspace{-3mm}
\caption[figure1]{\label{fig:drop} Quantum trajectory simulations of $^{39}$K incident on the optical potentials represented in figure \ref{fig:K_ITO} after falling from a height of 1mm.  The left (right) graph depicts 100 trajectories of the atoms falling towards ITO (ITO*) surfaces.  The highlighted green and red trajectories have the median reflected and absorbed escape times, respectively.  The ITO potential reflected 88\% of the atoms, while the ITO* potential reflected only 7\%.}
\end{center}\vspace{-0.3in}
\end{figure}

After falling 1mm, an atom initially at rest accelerates to .14m/s (for $^{39}$K, this corresponds to a kinetic energy of .96MHz).  Because the internal state of the $^{39}$K atom should relax at a rate of order 24.5$\mu$s$^{-1}$ (see section \ref{mirrors} and table \ref{vW_components}) and the near field enhancement feature depicted in figure \ref{fig:K_ITO} is much broader than the $\sim$10nm the atom would be expected to travel before reaching steady state, we can partially justify the use of the steady state approximation.  Similarly, the thermal de Broglie wavelength for a $1\mu$K (polarization gradient cooled \cite{CT89}) $^{39}$K atomic ensemble is only $h(2\pi m k_B T)^{-1/2}=7$nm, which supports the classical CM approximation for a typical cold atom ensemble.  On the other hand, as the laser fields in this example excite the atoms significantly (see figure \ref{fig:K_ITO}), incident atoms will heat from spontaneous emission scattering, complicating use of the steady state approximation.  To demonstrate the effects of spontaneous emission, illustrate some dynamics connecting the internal and external atomic variables, and numerically evaluate the steady state approximation introduced above, we consider quantum trajectory simulations \cite{Charmichael93} of $^{39}$K atoms falling from a height of 1mm onto the one-dimensional ITO and ITO* optical potentials represented by figure \ref{fig:K_ITO}.

Rather than describing the state evolution of a single atom, it is more accurate to interpret the atomic master equation \eqref{OBE} as evolving the mean internal state of an ensemble of independent atoms, without any measurements being made on the system \cite{Bouten07}.  This ``ensemble average'' evolution is ill-suited to demonstrate the dynamics between an individual atomic state and the forces on the atomic mass.  In contrast to its associated master equation, an ensemble of \emph{quantum trajectories} presume that at least partial measurements are being made on the system, which add a stochastic element to the dynamics, but helps preserve the state purity of each trajectory \cite{Charmichael93}.  In this subsection, we construct such trajectories by simulating experiments with photon counting on all radiation fields with perfect efficiency, spectroscopically filtered with a bandwidth small enough to perfectly distinguish spontaneous emission from the two atomic excited states, but much larger than any of the other dynamical rates in Eq. \eqref{OBE}.  Although not experimentally realistic, these trajectory simulations in particular help elucidate the effects of spontaneous emission on the motional dynamics of the atoms.  As purity is preserved in these simulations, the (unnormalized) evolution of a pure, internal atomic state is guided by the stochastic Schr\"{o}dinger equation ($\hbar=1$) \cite{vanHandel05}
\begin{equation}\label{SSE}
d|\psi\rangle = \left\{-i Hdt +\sum_{a,n}((\sigma_{na}-1)dQ_{na}-\frac12R^{tot}_{na}\sigma_{na}^\dag\sigma_{na}dt)\right\}|\psi\rangle
\end{equation}
where $|\psi\rangle$ is the internal state \emph{vector} of the atom and $dQ_{na}=1$ in each time step with probability $R^{tot}_{na}\langle\psi|\sigma_{na}^\dag\sigma_{na}|\psi\rangle dt$ and is zero otherwise.  And as above, we use a semiclassical approximation that the evolution of the (classical) atomic center of mass is driven by the force $\langle\psi|F(r_{CM})|\psi\rangle$, that spontaneous emission events ($dQ_{na}=1$) impart a random momentum kick with a probability distribution associated with the photon scattering randomly into $4\pi$ steradians, and that $r_{CM}$ determines the dynamical rates in Eq. \eqref{SSE} according to the atom-surface separation.

The K-ITO and -ITO* separation as a function of time in 100 quantum trajectories each is plotted in figure \ref{fig:drop}.  Despite an average 114 spontaneous emission events per drop, the ITO potential reflects 88\% of incident atoms with a turning point that suggests the steady state effective potential of figure \ref{fig:K_ITO}.  Similarly, the ITO* potential reflects a minority (7\%) of incident atoms, as to be expected from the initial kinetic energy (.96MHz) and figure \ref{fig:K_ITO}.  The imperfect reflection or transmission of atoms in the two cases is due to ``heating'' from non-deterministic scattering events.

\begin{figure}
\begin{center}
\includegraphics[width=4.5in]{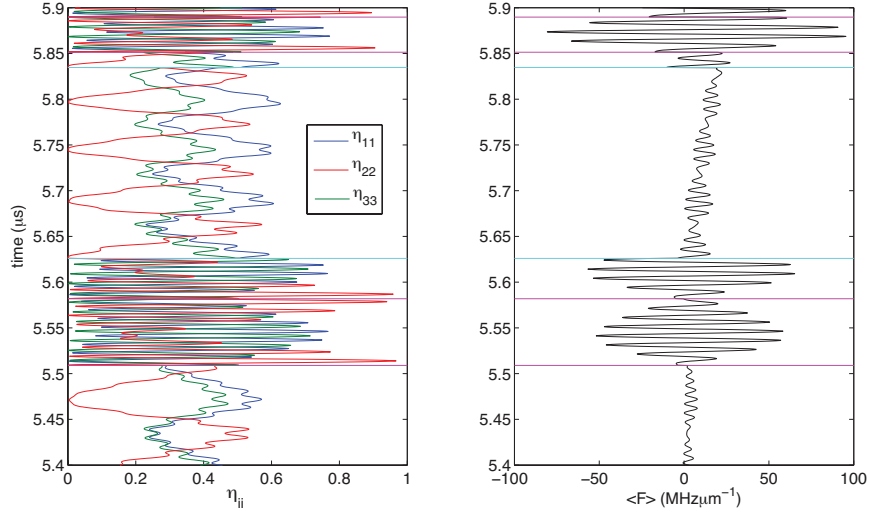}
%\vspace{-3mm}
\caption[figure1]{\label{fig:Force_close} A .5$\mu$s slice of a representative ITO trajectory.  Left: the instantaneous force eigenstate populations $\eta_{ii}$ as a function of time.  Right: the expected force on the atom as a function of time.  Purple and cyan lines indicate different types of photon counting events in the trajectory.}
\end{center}\vspace{-0.3in}
\end{figure}

Figure \ref{fig:Force_close} reveals some of the essential dynamics driving the trajectories, relating the internal atomic dynamics and the forces on the atomic center of mass.  A .5$\mu$s time slice of a representative ITO trajectory is given in both panels (during which the atom-surface separation was $\sim$200nm).  As a function of time, the (instantaneous position-dependent) force eigenstate populations $\eta_{ii}$ are presented in the left figure.  Purple (cyan) lines represent 1.178$\mu$m (767nm) spontaneous emission events, which instantly project the $^{39}$K atom into the $4P_{3/2}$ ($4S_{1/2}$) state.  The right figure depicts the force expectations $\langle\psi|F|\psi\rangle$ over the same time interval.  Because our particular choice of measurement projects the system into an atomic eigenstate, rather than an energy eigenstate, the internal states $\eta_{ii}$ undergo rapid and (roughly) coherent evolution between spontaneous emission events, giving rise to oscillations in the expected force on the atoms.  Although the exact evolution is specific to the system, separation, and choice of measurements in the simulation, it is important to note how each photon count abruptly ends the previous form of Schr\"{o}dinger-like evolution and initiates a new form, characteristic of the type of count that occurred.  Thus we see that in addition to the random momentum kick from photon scattering, spontaneous emission also gives rise to fluctuations in the dipole force that, although averaged over in the steady state potential Eq. \eqref{Ueff}, should be considered another source of heating.  However, this insight also provides a clear picture of how spontaneous emission rates can effect the \emph{time-averaged forces} on the atom: the total, time-averaged force may be estimated from the relative frequency of the spontaneous emission events and the time-averaged force of the Schr\"{o}dinger-like evolution they initiate (over a characteristic time interval).

\begin{figure}
\begin{center}
\includegraphics[width=4in]{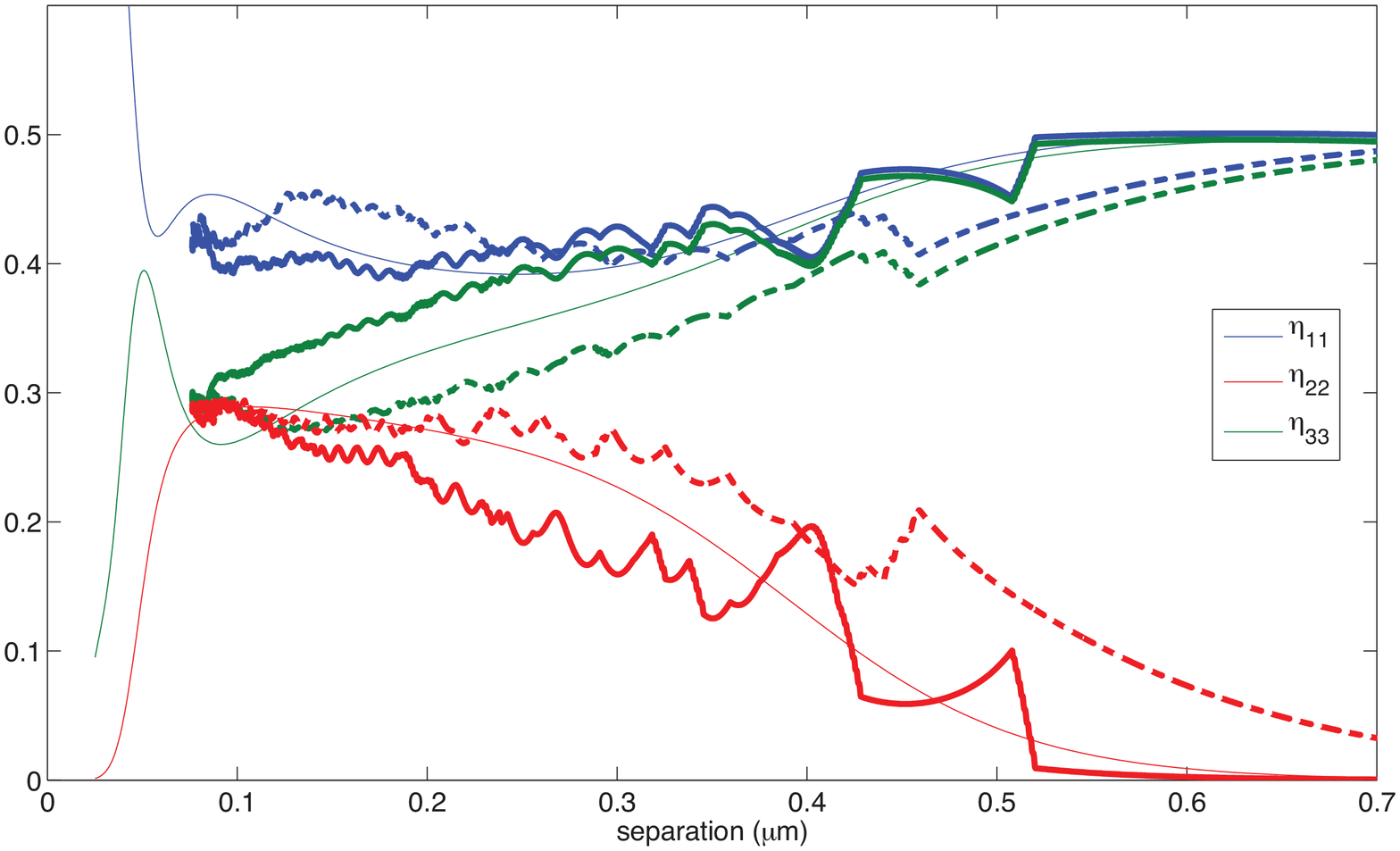}
%\vspace{-3mm}
\caption[figure1]{\label{fig:ss_vs_traj} Thick lines: quasi-mean $\eta_{ii}$ from a single quantum trajectory as a function of separation.  Solid (dashed) lines indicate the incident (reflected) portions of the trajectory.  Thin lines: steady-state $\eta_{ii}$ from figure \ref{fig:Force_ss}.}
\end{center}\vspace{-0.3in}
\end{figure}

Considering whole trajectories again, we re-examine the use of steady state approximations.  Thick lines in figure \ref{fig:ss_vs_traj} depict time-filtered $\eta_{ii}$ as a function of separation in a representative trajectory.  Low-pass, post-filtered with a cutoff frequency roughly 1/10$^{\text{th}}$ the free space spontaneous emission rates, these $\eta_{ii}$ represent the quasi-mean atomic populations relevant to the transient, atom mirror interaction.  While both incident and reflecting, the quasi-mean $\eta_{ii}$ follow the underlayed, steady state $\eta_{ii}$ from figure \ref{fig:Force_ss} (thin lines) with $\sim$.1 precision.  While figure \ref{fig:drop} suggested the validity of $U_{eff}$ at a single position (the classical turning point) over many trajectories, figure \ref{fig:ss_vs_traj} suggests the steady state potential for a single trajectory over a wide range of separations.

\begin{figure}
\begin{center}
\includegraphics[width=4.5in]{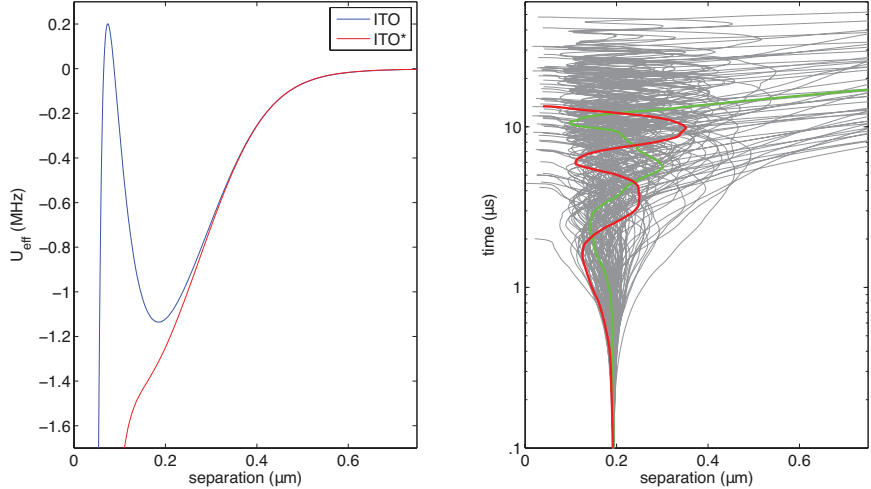}
%\vspace{-3mm}
\caption[figure1]{\label{fig:trap} Left: effective, steady state optical potentials with $\{\Omega_1,\Delta_1,\kappa_1,\Omega_2,\Delta_2\}/2\pi = \{100\text{MHz},50\text{MHz},(767\text{nm})^{-1},100\text{MHz},-3\text{MHz}\}$ that exhibit a near field potential minimum.  Right: 100 quantum trajectory simulations for $^{39}$K atoms initially at rest in the $^{39}$K-ITO potential minimum.  The highlighted green and red trajectories have the median reflected and absorbed escape times, respectively.}
\end{center}\vspace{-0.3in}
\end{figure}

Finally, we note that other useful, van der Waals-enhanced optical potentials may also be constructed.  For example, for the same optical parameters as in figure \ref{fig:K_ITO}, but with an additional, -3MHz detuning of the  $3D_{5/2}\leftrightarrow4P_{3/2}$ transition, a signifiant $U_{eff}$ minimum forms in the near field, as depicted in figure \ref{fig:trap}.  Quantum trajectory simulations of atoms initially at rest at the potential minimum suggest a mean trapping time of 20$\mu$s before heating effects eject the atom (figure \eqref{fig:trap}), whereas ITO* potentials produce no such trapping and lose atoms within $\mu$s'.  Although these trapping times are modest in this (non-optimized) potential, they are long compared to many fast atomic processes that characterize near-field atomic systems (e.g. \cite{Lev04}).  Traps and barriers with significantly lower heating rates would be possible for systems with anomalous van der Waals interactions that are an order of magnitude or more larger than the ground state perturbations (e.g. effects on the order of demonstrated in Cs and sapphire interactions \cite{Failache03}).

\section{Conclusion}\label{conclusion}

Although encouraging that significant surface-repulsion enhancement appears experimentally plausible with anomalous van der Waals effects of only moderate strength (predicted using published atomic and material data), many of the specific results of the previous section are of course limited to the case of $^{39}$K and carefully prepared ITO surfaces~\cite{Rhodes08} (although, in addition to the surface resonance emphasized here, the unique optical and electrical properties of ITO suggest a potential utility in integrated atomic systems in general). Although attractive pairings of alkali atoms and common materials are limited by an apparent dearth of high quality surface polariton resonances in the near infrared~\cite{Saltiel06}, engineering useful resonances with even simple surface patterning is a distinct possibility.  Guided in particular by Eq.~\eqref{G}, it is straightforward to recognize and analyze how construction of a transfer region (figure~\ref{fig:dipole_interface}) with planar thin films~\cite{Economou69}, gratings \cite{Barnes96,Pendry04}, or even material dopants \cite{Marquier04} could be used to perturb the dispersion of surface excitations, tuning them to useful atomic transitions. This line of research would naturally compliment current investigations that leverage similar effects using plasmonics to enhance and collect the spontaneous emission from embedded dipoles~\cite{Jun08,Akimov07}.

Harnessing surface effects for strong, near field manipulation of gas-phase atoms would enable integrated atomic systems and (at least in this implementation) employ an interesting intersection of fundamental QED, atomic physics, material science and surface polaritonics. Although exceptional, anomalous van der Waals effects admit an intuitive physical picture (section~\ref{interactions}) that should facilitate their generalization. And as suggested by section~\ref{example}, the effects need not be large for significant enhancement of near field potentials, and efficient steady state analysis may be reasonably accurate. However, we emphasize again that the identification of a broad class of surfaces with resonant excitations tunable to atomic transitions would open the door to general application of van der Waals-enhanced optical dipole potentials for atoms.

\section*{Appendix}
\appendix
\section{Calculation of van der Waals effects}\label{calculation}

Estimating the influence of a surface on atomic states is complicated, even in the near field where the field susceptibilities may be approximated by Eq. \eqref{G_nf}.  In this limit, with a simple planar interface, the energy shifts and spontaneous emission enhancement are
\begin{eqnarray}
\delta E_a &=& -\sum_n\frac{|\mu^{an}|^2+|\mu^{an}_z|^2}{16z^3}\left(\frac2\pi \int_0^\infty d\zeta\frac{\epsilon_1(i\zeta)-1}{\epsilon_1(i\zeta)+1}\frac{\omega_{na}}{\omega_{na}^2+\zeta^2}+2\text{Re}\frac{\epsilon_1(\omega_{an})-1}{\epsilon_1(\omega_{an})+1}\Theta(\omega_{an})\right)\nonumber\\
&\equiv&-\sum_nM_{an}(\Delta_{na}^{vf}+\Delta_{na}^r)z^{-3}\label{dE_nf}\\
R_{na} &=& \frac{|\mu^{an}|^2+|\mu^{an}_z|^2}{16\hbar z^3}4\text{Im}\frac{\epsilon_1(\omega_{an})-1}{\epsilon_1(\omega_{an})+1}\Theta(\omega_{an}) \equiv \hbar^{-1}M_{an}r_{na}z^{-3}\label{R_nf}.
\end{eqnarray}
We can therefore break up our calculation into separate parts: the dipole transition strengths, $M_{an}$, and the image factors, $\Delta_{na}$ and $r_{na}$.

We will neglect any atomic hyperfine structure and assume an isotropically prepared atomic dipole for calculation and conceptual clarity.  The widths of $\epsilon(\omega)$ features will be on the order of .1eV, while typical alkali hyperfine splittings are only of order 10GHz or less \cite{Arimondo77}.  Thus image factors should not vary significantly between hyperfine states.  However, these splittings are of order the energy scales of typical laser Rabi frequencies and detunings.  Moreover, atomic dipole moments vary greatly between different hyperfine spin projections.  A more thorough analysis of $M_{an}$ should be considered in any experimental implementation.

Under these approximations, we may estimate the atomic dipole strength with \cite{Chevrollier92}
\begin{equation}\label{dipole_strength}
|\mu^{an}|^2+|\mu^{an}_z|^2 = \hbar R_{\{n,a\}}^{fs}\left(\frac{c}{|\omega_{an}|}\right)^3(1+2\frac{J_n-J_a}{2J_a+1}\Theta(\omega_{na}))
\end{equation}
where $R_{\{n,a\}}^{fs}$ is the free space decay rate for either $a\rightarrow n$ or $a\leftarrow n$, and $J_i$ is the spin-orbit quantum number of state $i$.  Using atomic data from \cite{Corliss79} and \cite{Heavens61}, we list the calculated $M_{an}$ for the $4S_{1/2}$, $4P_{3/2}$, and $3D_{5/2}$ levels of neutral $^{39}K$ in table \ref{vW_components}.

The image factors may likewise be calculated from the bulk material's dielectric function.  In modeling the response of indium tin oxide (ITO), we follow  \cite{Rhodes08} and use an empirical Drude free-electron model
\begin{equation}\label{Drude_ITO}
\epsilon_1(\omega) = \epsilon_\infty - \frac{\omega_p^2}{\omega(\omega+i\gamma)}
\end{equation}
with measured material constants $\epsilon_\infty = 3.8$, $\omega_p = 2.19$eV, and $\gamma = .111$eV.  Using Eq. \eqref{Drude_ITO}, we list the calculated $\Delta_{na}$ and $r_{na}$ in table \ref{vW_components}.

\begin{table*}
\caption{\label{vW_components} Calculation of the transition dipole strengths and image factors for the  $4S_{1/2}$, $4P_{3/2}$, and $3D_{5/2}$ levels of neutral $^{39}K$ and ITO.  Data taken from \cite{Corliss79,Heavens61,Rhodes08}.}

\begin{tabular}{cc|cc|c|ccc}
\hline
$a$ & $n$ & $\omega_{na}\text{(eV)}$ & $R_{\{n,a\}}^{fs}(\mu s^{-1})$&  $M_{an}(\text{kHz}\mu\text{m}^3)$ &  $\Delta_{na}^{vf}$ & $\Delta_{na}^r$ & $r_{na}$ \\ \hline\hline
$4S_{1/2}$ & $4P_{1/2}$ & 1.6093 & 36.9 & .6758 & .7373 & & \\
 & $4P_{3/2}$ & 1.6165 & 37.2 & 1.3447 & .7368 & & \\
 & $5P_{1/2}$ & 3.0613 & 1.98 & .0053 & .6821 & & \\
 & $5P_{3/2}$ & 3.0637 & 2.14 & .0114 & .6821 & & \\ \hline
 $4P_{3/2}$& $4S_{1/2}$ & -1.6165 & 37.2 & .6723 & -.7368 & .6570 & .1133 \\
  & $5S_{1/2}$ & .9894 & 14.2 & .5595 & .7856 & & \\
  & $3D_{3/2}$ & 1.0527 & 4.09 & .2627 & .7793 & & \\
  & $3D_{5/2}$ & 1.0524 & 24.5 & 2.4067 & .7794 & & \\ \hline
$3D_{5/2}$ & $4P_{3/2}$ & -1.0524 & 24.5 & 1.604 & -.7794 & -2.3873 & 7.6597 \\
& $5P_{3/2}$ & .3948 & 1.58 & 1.3068 & .8741 & & \\
\hline
\end{tabular}
\end{table*}

\begin{table}
\centering\caption{\label{vW_table} Total van der Waals energy shifts and enhanced decay rates for $^{39}$K, ITO, and ITO* as determined by Eqs. \eqref{dE_nf} and \eqref{R_nf}, and table \ref{vW_components}.}

\begin{tabular}{r c l}
\hline
$\delta E_a$,$R_{na}$ & ITO (kHz$\mu$m$^3$) & ITO* (kHz$\mu$m$^3$)\\ \hline\hline
$\delta E_{4S_{1/2}}$ & -1.5 & -1.2\\
$\delta E_{4P_{3/2}}$ & -2.3 & -2.3\\
$\delta E_{3D_{5/2}}$ & 3.9 & -1.7\\ \hline
$R_{4S_{1/2}\leftarrow4P_{3/2}}$ & $2\pi\times.76$ & 0\\
$R_{4P_{3/2}\leftarrow3D_{5/2}}$ & $2\pi\times12.3$ & 0\\
\hline
 \end{tabular}
 \end{table}

The aggregate van der Waals effect for each atomic level is determined by Eqs. \eqref{dE_nf} and \eqref{R_nf}, and the entries in table \ref{vW_components}.  The results are presented in table \ref{vW_table} in comparison to a dispersionless dielectric called ``ITO*'' with $\epsilon_1(\omega) = \epsilon_\infty$.  The largest effect is the enhanced decay $R_{4P_{3/2}\leftarrow3D_{5/2}}$.  The nearby surface polariton resonance in ITO ($\epsilon_1(1.0524\text{eV}) = -0.4826 + 0.4517i$) \emph{and} the relatively strong $3D_{5/2}\rightarrow4P_{3/2}$ dipole moment (see table \ref{vW_components}) combine for a \emph{net positive energy shift} and a significantly enhanced spontaneous emission rate.

\section*{Acknowledgments} This work has been supported by an HP Labs Innovation Research Award and by the NSF under CCF-0622246. JK thanks Young Chul Jun, Jon Schuller and Mark Brongersma for many useful discussions.

\end{document}